# Extended habitability of exoplanets due to subglacial water

Amri Wandel
Racah Institute of Physics, The Hebrew University of Jerusalem, Jerusalem, Israel

## Abstract

Considering subglacial liquid water, a significant extension of the classical Habitable Zone is obtained. Elaborating on the model of Wandel (2023) it is shown how an atmosphere and liquid water could survive on tidally locked planets closely orbiting an M-dwarf host, extending the Habitable Zone boundary inwards. In addition, subglacial liquid water could extend the Habitable Zone beyond the outer boundary of the conservative Habitable Zone as well. These two results enhance the circumstellar region with a potential for liquid water well beyond the conservative boundaries of the classical Habitable Zone. It is argued that the probable recent JWST detection of atmospheric water vapor on the rocky Earth-sized exoplanet GJ 486 b, along with earlier detections of water on other planets orbiting M-dwarf stars gives an empirical answer to the much argued question, whether such planets can support liquid water, organic chemistry and eventually life. It is shown how water on terrestrial planets closely orbiting M-dwarf stars may sustain in a subglacial melting layer. Finally, the model is applied to a few exoplanets demonstrating how the detection of water may constrain their atmospheric properties.

Key words: planets: climate, planets: atmosphere, Habitable Zone, M-dwarfs, tidal locking, basal melting, subglacial water

## 1. Extending the Habitable Zone

The capability to maintain liquid water on a planetary surface is considered as a condition for habitability, defining the boundaries of the Habitable Zone (HZ). The classical HZ is the circumstellar region where a planet's surface temperature allows for liquid water to exist, considering the stellar irradiation, or instellation, and the planetary atmospheric conditions. The inner boundary of the habitable zone is determined by the point at which water would evaporate due to excessive heat even before the boiling point is reached, due to processes such as the moist greenhouse effect (Kasting et al. 1993), while the outer boundary is determined by water freezing due to insufficient heat. These boundaries are consistent with the 1D climate models (Korappapu et al. 2013; Yang et al. 2014a). Alternatively, somewhat wider empirical HZ boundaries may be drawn by assuming that early Mars (outer boundary) and recent Venus (inner boundary) had liquid water. These HZ-boundaries have a weak dependence on the stellar type (Shields 2016) due to the difference in wavelength of the stellar irradiation. Fig. 1 illustrates the HZ boundaries given by these models as a function of the instellation and the star type, which is denoted on the vertical axis by the stellar surface temperature.

Planets in the HZ of M-dwarf stars are likely to be tidally locked (Shields et al. 2016) and may have temperatures adequate for liquid water, at least on part of their surface, under a broad range of atmospheric conditions (Wandel 2018). However, liquid water may exist out of the classical HZ due to geothermal heat as a subglacial ocean (e.g. in Europa and Enceladus, due to tidal heating by Jupiter and Saturn, respectively). Alternatively, geothermal heat may come from radiogenic elements, as calculated by Ojha et al. (2022). They modeled the evolution of ice sheets and the conditions that allow liquid water to be produced and maintained at temperatures above freezing on Earth-like exoplanets. Applying their results Wandel (2023) enhanced the definition of habitability by including subglacial liquid water, suggesting a more extended circumstellar habitable region with a potential for liquid water. In colder planets, outside the traditional HZ geothermal heat may suffice to maintain basal melting. On the other end, planets that are closer to their host star than the inner boundary of the conservative HZ, which were until now considered too hot, may also, under certain conditions, support subglacial or liquid water, **not** due to geothermal heat but rather being protected from the stellar irradiation under ice sheets that accumulate on the night side of tidally locked planets (section 3). As planets of M-dwarfs are the most readily observable for biosignatures (Shields et al. 2016), subglacial oceans eventually harboring organics and possibly life may be more easily detected by spectral transmission analyses, assuming that subglacial water can reach the surface producing geysers or plumes, as observed in Europa and Enceladus (e.g. Huybrights et al 2020; Barge and Rodriguez 2021).

In this work we extend the model of Wandel (2023) and apply it to recent observations, obtaining two important new results: (i) the detection of water vapor in GJ 486 b is used as an empirical support for the capability of closely orbiting rocky planets of M-stars to maintain or renew water and atmosphere, contrary to previous suggestions that stellar activity during the early evolutionary stage of M-dwarfs erode the atmosphere and destroy eventual water on such planets (section 4), and (ii), the extended HZ boundaries due to subglacial water are combined with the analytical climate model of Wandel (2018) to demonstrate that habitable surface temperatures are consistent with a fairly wide range of atmospheric properties for several representative well known HZ Earth- and Superearth-size planets of M-stars (section 6).

## 2. Basal melting and subglacial liquid water outside the HZ

The HZ extends between the inner boundary (water evaporates) and the outer one (water freezes). At a larger distance from the host star, too low temperatures do not allow the existence of liquid water on a planet's surface. Basal melting of local or global ice sheets by geothermal heat may in that case be an alternative to radiative heat from the host star. Ojha et al. (2022) have shown that subglacial liquid water may be maintained even on planets with modest geothermal heat (as low as 0.1 Earth's, for sufficiently thick ice layers), produced by radiogenic elements. They find that subglacial oceans or lakes of liquid water can form by basal melting and persist under ice sheets on Earth-sized exoplanets even for surface temperatures as low as 200K. Under the weight of high-pressure ices, the basal meltwater is buoyant and can migrate upward, feeding intra-glacial lakes and eventually a massive water layer or ocean, possibly in contact with the silicate base, like in Titan and Ganymede (Kalusanova and Sotin 2018). Basal melting is more likely to occur on planets with thicker ice sheets, higher surface gravity and higher surface temperatures.

As a conservative outer limit for the HZ extended by basal melting, one may use the flux on Trappist 1g, the coldest exoplanet calculated by Ojha et al· (2022), which receives about 25% of the radiative flux received by Earth (hereafter denoted by $S_E$). This value is roughly coincident with the outer boundary set by moist greenhouse limit in the 1D model (Kasting et al 1993; Yang et al 2014a), as well as with the Early Mars hypothesis (an empirical HZ boundary derived from the flux received by Mars a few billion years ago, considering the evidence from the Mars missions, that the early Mars had abundant surface liquid water).

We suggest a lower flux boundary, that is a more extended outer boundary of the HZ, derived from the evidence for an intra-glacial lake in Mars's south pole region (Orosei et al. 2018). Estimating the annual average of radiative flux received by Mars' polar region as *<cos(i)>~0.3* of the irradiation at Mars' orbit (*i* being the angle of incident radiation) gives a flux of approximately *0.1 $S_E$*, which is denoted in Fig. 1 by the violet dashed line marked "Martian polar lakes". The weak dependence of the flux boundaries on the stellar type is assumed to be similar to the dependence of the moist greenhouse boundary.

In addition to radiogenic heat, subglacial liquid water may be produced by tidal heating. This mechanism can maintain basal melting on planets which do not have enough geothermal radioactive heat production, e.g. small or old planets, if they are close enough to their host star, or on moons on tight orbits around giant planets. The former case may be particularly relevant to small (Earth-sized and smaller), as such planets are likely to cool geologically within a few billion years, and planets closely orbiting M-dwarf stars may be tidally heated.

Like the conservative HZ boundaries, also the extended boundaries are not constant in time but change slowly due to the evolution of the host star (in the case of instellation-driven HZ) and due to the evolution of the internal planetary thermal structure (in the case of geothermally controlled HZ). In particular, the outer boundary may expand in the case of intraglacial lakes, as the luminosity of the host star increases due to evolution on the Main Sequence. On the other hand, if the outer HZ-boundary is determined by geothermal heat, it may get closer to the star over the geological timescale of the planet, as radiogenic heat diminishes. In the case of closely orbiting planets around M-stars this shrinking stops when the geothermal heat reaches the level of tidal heating.

## 3. Inwards of the Habitable Zone

At a distance closer to the host star than the inner boundary of the conservative HZ (surface temperature too high) processes such as the wet greenhouse effect or greenhouse runaway may lead to rapid evaporation of surface liquid water. Planets orbiting close enough to the host star (in particular planets in the HZ of M-dwarfs and closer) are tidally locked, with one side permanently facing the host star, or on a synchronous orbit (Heller et al. 2011). On tidally locked or slowly rotating planets, greenhouse runaway evaporation may be avoided as low surface temperatures may be maintained on the night side (Wandel and Gale 2020).

Rocky planets with little or no atmosphere (like Mercury) may have large temperature differences between the day and night sides, unless heat is efficiently transported from the day side, e.g. by atmospheric convection (Wandel 2018). While the day side may be too hot for maintaining liquid water, the water on the night side may accumulate as ice (Leconte et al. 2013), unless heat transport is sufficient to evaporate the ice on the night side. If the heat transport is low, ice can form on the night side and subglacial liquid water may provide a potentially habitable neighborhood.

Calculations with 3D Global Climate Models suggest an atmospheric collapse scenario: water evaporating at the sub stellar side, transported by atmospheric currents to the night side where it is trapped as ice caps (Leconte et al. 2013; Yang et al. 2014b). Basal melting may then be driven by radiogenic elements as well as by tidal geothermal heating due to the gravity of the host star.

According to the arguments above, taking into account subglacial liquid water the classical HZ may be extended inwards. The inner boundary of the extended HZ would occur at the point where the heat transported from the day-side of the locked planet is sufficient to melt and eventually evaporate the ice on the night side, or prevent its formation in the first place. This boundary would depend on the fraction of the heat transported from the sub stellar day side to the night side. Following Wandel (2018) we define the heat transport parameter $f$, which is the fraction of irradiated energy at any given latitude which is transported (e.g. by atmospheric currents or jet streams), assumed to be evenly distributed over the whole planet.

As reference for the heat sufficient for night-ice evaporation we take the radiative flux of the moist greenhouse limit (Kasting et al. 1993). Fig. 1 shows three inner boundaries (blue dashed curves), for $f=0.3, 0.1$ and $0.03$ (marked 30%, 10% and 3% respectively). The lower the transported fraction $f$, the closer to the host star is the inner extended HZ boundary.

We find that for $f<0.7$ the night side basal melting boundary is closer to the host than the optimistic inner HZ boundary according to the Recent Venus hypothesis (an empirical value derived from the flux received relatively recently by Venus, assuming that until recently it had liquid water). As argued above, ice may exist that close to the host star only **on the night side of** tidally locked planets.

For comparing the HZ and tidal boundaries we derive an expression for the tidal locking radius $r_{tl}$ in terms of the radiative flux from the host star. Using the tidal locking relation (Burns and Matthews 1986; Gladman et al. 1996) $r_{tl} \sim M^{1/3}$ and the approximate Main-Sequence relations between stellar mass $M$, luminosity $L$, and surface temperature, $L \sim M^{3.5}$ and $T \sim M^{1/2}$ (assuming the host is a Main Sequence star) we relate it to the stellar type. For rocky planets (normalizing to the parameters of Mercury) we find the approximate relation

$$S(r_{tl})/S_E \sim 7.3 \ (T/T_\odot)^{5.7},$$

where $T_\odot$ is the surface temperature of the sun. This is shown in Fig. 1 as the orange curve, the width of which represents the uncertainty in the above analyses.

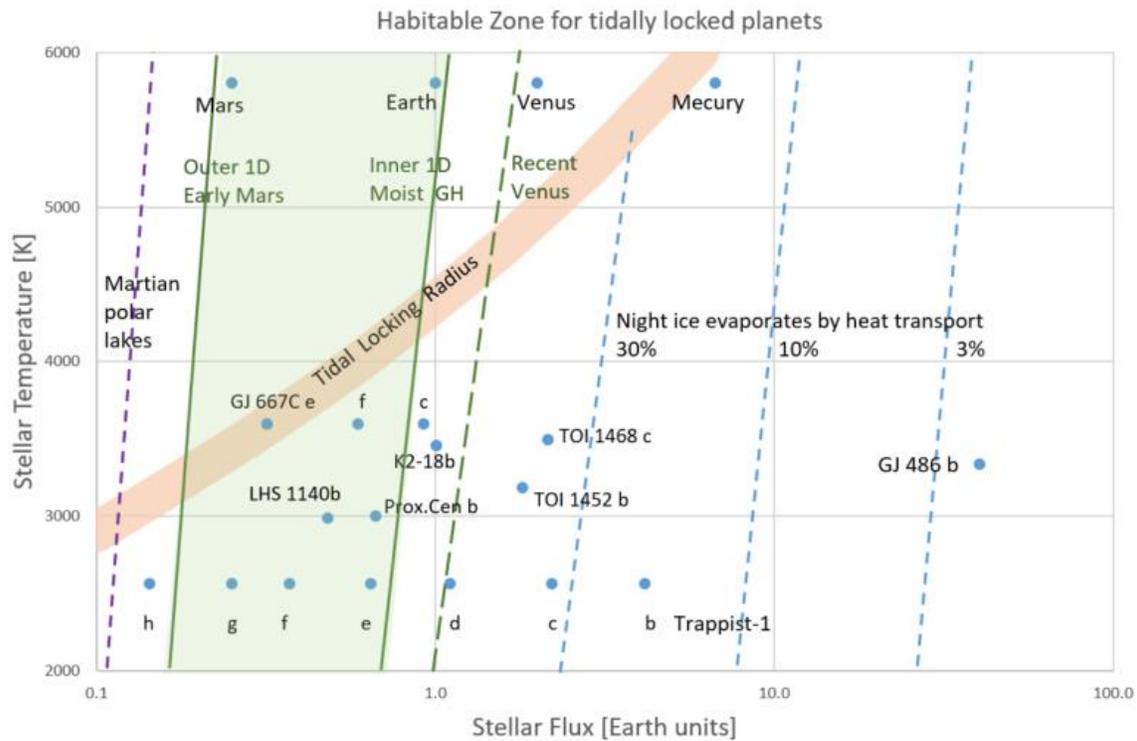

Figure 1: Boundaries of the Habitable Zone under various climate and atmospheric models as a function of radiative flux received from the host star, relative to Earth. Dashed lines represent the extended boundaries due to subglacial liquid water suggested in this work: blue for the inner HZ boundary (night-side ice evaporates due to heat transport, the percentage marks the global transport parameter $f$) and violet for the outer one (Martian polar lakes). Solid green lines and green shading denote the conservative HZ boundaries: inner one (moist greenhouse) and outer one (maximum $CO_2$ greenhouse), calculated using the 1-D climate model. Also marked is the Recent Venus boundary (dashed green line). The orange curve marks the tidal locking radius. Circles denote the terrestrial planets in the solar system and a few Earth-sized exoplanets orbiting M-dwarf stars.

## 4. Can an atmosphere and water survive on planets of M-dwarfs?

It is often stated that planets orbiting M-stars may be effected by the host's energetic outbursts in the early evolutionary stages, producing intense UV and X-ray radiation and stellar eruptions that may disintegrate water molecules (Luger and Barnes 2015). While super-Earth and larger planets are likely to have a thick atmosphere that may survive the early active evolutionary phase of M-dwarfs, as may be the case of 2K-18b, demonstrated by the detection of water vapor in its atmosphere (Tsiaras et al. 2019), the survival of water and atmosphere on smaller, Earth-sized planets orbiting M-dwarf stars seemed less certain.

The recent detection of water vapor on the rocky Earth-sized exoplanet GJ 486 b (Moran et al 2023), if associated with the planet rather than with cool star spots, may provide an empirical support to the sustainability of water and atmosphere also on small planets of M-dwarfs. Even if the primary atmosphere was eroded, water may have survived in subglacial layers and eventually, after the end of the host's early evolutionary phase, a secondary atmosphere may be produced by volcanic activity and enriched by water vapor reaching the surface through cracks or fissures in the ice.

The age of GJ 486 b is estimated between 1-8 Gyr (Caballero, et al. 2022). Its estimated Main Sequence lifetime is >30 Gyr, so that GJ 486 b could still be close to its active period. Also, with an irradiation flux of $40 S_E$ and a calculated equilibrium temperature of ~700K, water vapor in the atmosphere may have been destroyed as on Venus. Hence the detection of significant amounts of water vapor in the atmosphere of GJ 486 b could indeed serve as an empirical support for the model of subglacial night side water on tidally locked planets of M-dwarfs.

In addition to the intensive energetic UV and X-radiation from the young M-dwarf host, liquid water and habitability of planets of M dwarfs may be threatened by atmospheric erosion by stellar winds (Rodríguez-Mozos and Moya 2019). Isolated from space by the ice cover, subglacial oceans can survive atmospheric erosion and exist even without or with only a thin atmosphere. Protected from energetic UV and X-ray radiation from the young M-dwarf host, such lakes or oceans could provide a safe neighborhood for the evolution of organic chemistry.

## 5. Abundance of potentially biotic exoplanets

Subglacial liquid water may exist out of the traditional HZ also on planets of K- and G-type stars, but the most significant contribution to the abundance of habitable planets comes from M-dwarfs. Being the most abundant stellar type, extending the HZ around M stars could have a major effect on the abundance of potentially habitable planets, i.e. small rocky planets with a potential to support liquid water and biochemistry on the surface or under an ice cover. The extended habitability region described above, in particular around the abundant M-dwarfs, may significantly enhance the number of biotic planets. Wandel (2023) estimated that considering planets of M-dwarfs and an extended HZ may increase the abundance of potentially habitable worlds by a factor of up to ~100, compared to only G-type stars with a conservative HZ. This estimate is based on the known relative abundances of M, K and G-type stars and the Kepler statistics of Earth-size planets in the HZ depending on the star type (Bryson et al. 2020), combined with the enhancement of the HZ. While G-type stars are just 4% of all stars, K-type stars are ~16% and M-dwarfs consist 72% (Glovin et al. 2023). Hence if K- and M-dwarfs could support liquid water, the abundance of potentially habitable planets would increase ~20-fold, compared with the conservative assumption that only G-type stars can. The Kepler mission found that between 10%–50% of all stars host a small, rocky planet within their habitable zone (Cassan 2012; Dressing and Charbonneau 2015). The reason for this wide range is the uncertainty of the climate model, which determines the HZ boundaries (the green boundaries in Fig. 1). Hence compared to a conservative HZ the extended HZ would enhance the number of potentially habitable planets by at least another factor of 5, which together with the above factor of 20 due to stellar population gives a factor of 100. Figure 2 of Wandel (2023) shows the number of expected habitable planets as a function of the distance from Earth for various star types and HZ models. For example, within

30 pc from Earth about 100 habitable planets are expected with the most conservative assumptions (only G-type stars and a classic HZ) while for the most optimistic assumptions (including M-type stars and an extended HZ) their number could be as high as 10,000.

## 6. Atmospheric properties and habitability

As shown in section 3 and fig. 1, the extension of the HZ allowing for subglacial liquid water closer to the host star, or equivalently, to higher values of stellar radiative flux (instellation), depends on the amount of global heat transport, quantified by the parameter *f*. An additional important global property of the atmosphere, which determines the planetary surface temperature distribution is the is the atmospheric heating, defined by Wandel (2018). Intuitively it is the amount by which the atmosphere enhances the average surface temperature of the planet, *T*, compared to the equilibrium isothermal surface temperature of the planet $T_{eq}$, which is calculated without taking into account an atmosphere: $4\sigma T_{eq}^4=(1-A)S$, where $\sigma$ is the Boltzmann constant, *A* is the Albedo of the planet's surface, and S is the instellation. Using this definition, the atmospheric heating can be written as

$$H_{atm}=(T/T_{eq})^4.$$

For example, the atmospheric heating of the terrestrial planets of the solar system (Venus, Earth and Mars) are 26, 1.15 and 0.75 respectively. The atmospheric heating consists of three elements: the amount of greenhouse heating ($H_g$) combined with the Albedo and atmospheric screening $\alpha$:

$$H_{atm}=(1-A)\ \alpha\ H_g$$

In the analytic climate model of Wandel (2018) the surface temperature distribution is determined by three parameters: the global heat transport factor *f*, the atmospheric impact given by $H_{atm}$ and the instellation $s=S/S_E$ (where $S_E$=1361 W/m² is the energy flux received by Earth). In this model, introducing a specific temperature range (*0<T<100* C) gives the allowed range for the atmospheric heating as a function of the other two parameters:

$$0.23(1-3f/4)^{-1}< sH_{atm} <3.2\ f^{-1}$$

Of course, these boundaries for the surface temperature and do not directly apply to subglacial water, which being maintained by geothermal heat at the outer boundary of the HZ is near 0 C. Nevertheless, the above temperature range is relevant to the case of instellation-dominated HZ at the inner HZ boundary, via the requirement that heat transport would not evaporate the night side ice, as well as to the conservative outer HZ boundary. This is demonstrated in fig. 2, which shows the atmospheric heating ranges allowed for GJ 468 b and a few other representative planets, also marked in fig. 1. These ranges may be compared with the constraint on the heat transport required to avoid evaporation of the night side ice. The strongest constraint is for GJ 468 b, which requires $f<0.03$. The equation above gives an atmospheric heating $0.23<H_{atm}<\sim 4$. Although the constraint on the heat transport implies a quite thin atmosphere, there is still a fairly wide range for the atmospheric heating. For TOI 1468 c the constraint on the heat transport is much less severe, $f<0.4$, and a similar upper limit is inferred from fig. 2, $0.2<H_{atm}<4$. For the two other planets shown in fig. 2 there is no constraint on the heat transport. In particular, even $f\sim 1$ is allowed by the ice evaporation criterion, so a dense atmosphere is not excluded, consistently with their higher masses: K2-18 b is 8.6±1.3 times the Earth mass, and has probably a massive atmosphere (this is supported by the low mean density, 2.7±0.5 g/cm³ and the observed traces of hydrogen (Tsiaras et al. 2019). A heat transport of ~1 yields a somewhat narrower range for the atmospheric heating, $1<H_{atm}<3$. Also LHS 1140 b, with a mass of 6.4±0.4 times the Earth mass, may have a massive atmosphere with $f\sim 1$, giving $2<H_{atm}<7$.

These figures demonstrate that the model can consistently explain the observed planetary properties, in particular the water vapor on K2-18 b and GJ 468 b, with only modest constraints on their atmospheric parameters.

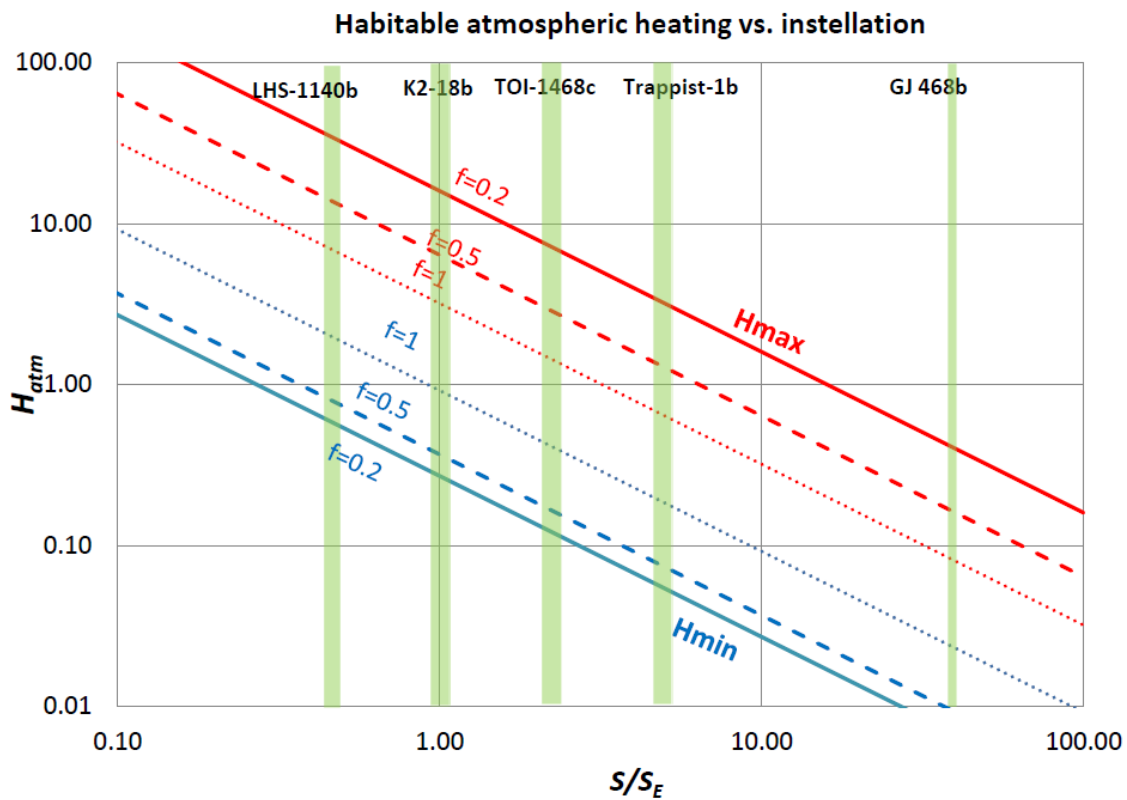

Fig. 2 Maximal (red) and minimal (blue) boundaries of the atmospheric heating factor vs. instellation, for three values of the heat transport parameter: $f=0.2$ (solid), 0.5 (dashed), and 1 (dotted). GJ 468 b and a few representative planets from fig. 1 are marked as green vertical bars, with the width showing the uncertainty in the instellation.

## 7. Conclusion

Basal melting and subglacial liquid water may provide habitable environments on the night side of closely orbiting locked planets of M-dwarfs. Such water would be shielded against the energetic radiation of the host star in its early evolutionary stages. Also on planets more distant from the host star, too cold for surface liquid water, basal melting may feed long lasting subglacial lakes outside the conservative HZ. The recent detection of water in the transmission spectrum of the rocky Earth-size planet GJ 486 b and the super Earth K2-18b can be explained by this model, with only modest constraints on their atmospheric parameters, supporting the sustainability of atmosphere and water on planets closely orbiting M-dwarf stars.